\documentstyle[12pt]{article}

\begin{document}

\begin{center}
{\Large \bf High Angular Resolution and Extrasolar Planets: Beyond
Basic Instrumental Performances} 

\bigskip

{\large Jean Schneider CNRS - Paris Observatory, 92195 Meudon, France
-- Jean.Schneider@obspm.fr}
\end{center}

\mbox{}\vspace{2cm}\\

{\large \bf Abstract}
 
I review the characteristics of planetary systems accessible to imaging. 
I show that, beyond the basic angular resolution and
dynamics performances of an optical ``architecture'', other 
performances such as photometric 
precision, spectral range or timing are necessary to access some physical 
characteristics of the
planets.\\

\bigskip

\section{Introduction}
  The main problem faced by the imaging of planetary systems is to separate
the planets from their parent star. This requires two equally important 
capabilities: a sufficent angular resolution to separate the planet from the 
star and a method of supression of stellar light to fight the very high
star to planet contrast (of the order of $10^6$ to $10^9$).
But many characteristics of the planetary system require other capabilities 
and the present paper is intended to point out that
the traditional
couple ``High angular resolution and contrast'' is not the final response 
to all the scientific questions.
\section{Characteristics of planets accessible to imaging}
{\bf Mass:}
The most traditional quantity accessible beyond radial velocity
measurements is the planet mass. The latter is measurable by astrometry.
The astrometric precision forbidds to measure mass below a few Eerth masses.
Another way to measure planet masses, with a much higher precision and down to
fractions of Mars masses is to detect satellites of light planets and
to infer the mass from the satellites revolution period and distance
to the planet from Kepler laws (Schneider and Riaud 2001). 

{\bf Radius $R_{pl}$:}
It is can in principle be deduced from the planet thermal flux $(F_{pl})_{th}$
given by
\begin{equation} 
(F_{pl})_{th} = 4\pi \sigma R_{pl}^2\times T_{pl}^4 
\end{equation} 
 The planet temperature $T_{pl}$ can be infered from the matching of a
planckian function to the observed thermal spectrum (forgetting here 
complications due to absorption features).

{\bf Albedo $A$:}
It can be measured from the reflected flux $(F_{pl})_{refl}$ given by
\begin{equation} 
(F_{pl})_{refl}(t) = F_* \times A\times \left(\frac{R_{pl}}{2a}\right)^2 
\times\phi(t)  
\end{equation} 
 where $\phi(t)$ is an orbital phase factor. For a circular orbit with an
``annual'' period $P$, it is given by
\begin{equation}
\phi(t) = (1 - \sin i \sin (2\pi t/P))/2
\end{equation}
The measurement of the albedo (generally in the visible) thus requires
the measurement of the thermal flux (generally in the mid-infraed region).
 
But this scheme faces some complications which are {\it a  priori} not
exceptional.
Indeed, formulae (2.1) to (2.3) are valid only for spherical bodies.
There are at least two situations for which a planetary companion cannot be 
identified to a single spherical body: planetary rings and ``binary planets''.

{\it Rings:}
In that case:

 (1) part of the planet can be hidden by the ring (and vice versa)

 (2) The ring being not spherical, for half of the orbit the observer sees
its back side which is not illuminated by the star and thus completely dark
(Schneider 1999 \cite{Schneider99}).
The planet + ring total flux can then have an orbital
dependence completely different from (2.3), even in case of circular orbit.
The solution is then to measure $\phi(t)$ along the orbit. That requires the 
capability
to revisit the target several times (more then 10 along the orbit).

{\it Binary planets:}
By ``binary planet'' I mean a planet with a large satellite companion which 
can eventually be as large as the planet itself. In the Solar System we do not 
know such binary planets. But since binary astero\"{\i}ds and brown dwarfs
haved been discovered these last years, it is not an unrealistic speculation
that binary planets may exists in other planetary systems. They would not 
 constitute just
a curiosity, their existence would disturbe the planet mass function 
deduced from dynamical (radial velocity, astrometric and timing)
measurements.

How the separate a single from a binary planet?
Even if the binary planet cannot be resolved, 
a straightforward way is to measure the position 
variations of the binary planet photocenter. 
Its amplitude $(\Delta x)_{ph}$ along a binary orbital period (``month'')
is given by
\begin{equation}
(\Delta x)_{ph} = \frac{|F_1a_1 - F_2a_2|}{F_1+F_2}
\end{equation}
where $F_{1,2}$  and $a_{1,2}$ are the flux and the distance to 
the center of mass of the binary planet for each component.
For the reflected stellar light and the thermal emission of the planets
(assuming a rotation rate for each component sufficiently large to 
ensure a uniform temperature),
$(\Delta x)_{ph}$ is  of the order of 1 to 10 $10^{-3}$ AU (0.1 to 1 mas at 
10 pc).
A special mention must be made for tidally locked satellites.
For that configuration, the satellite has, at any time, a cold and a warm face.
In addition to the photocenter motion, there is a variation in the thermal 
flux as seen by the observer along the annual orbit (Woolf 2001).
The corresponding annual variation of the effective temperature of the
 binary planet given by  
\begin{equation}
T_{eff} = (1+(R_1/R_2)^2)[T_{min}^4(1+\cos \alpha)/2+T_{max}^4
(1-\cos \alpha)/2])^{1/4}
\end{equation}
where $\alpha$ is the angle between the line of sight and the star to planet
direction.
The induced monthly  photocenter motion has an amplitude
\begin{equation}
(\Delta x)_{ph} = 2(a_1+a_2)\frac{(M_1T_1^4R_1^2-M_2T_2^4R_2^2)}
                     {(M_1+M_2)(T_1^4R_1^2+T_2^4R_2^2)}
\end{equation}
For instance, for an Earth with a large moon (half the Earth  size),
the photocenter amplitude is 3.3 10$^{-3}$ UA (=
0.3 mas 10 pc).\\

{\bf Variations of the planet temperature and albedo:}

{\it Annual:} 
It can be due to the eccentricity of the orbit or to the inclination
of the planet rotation axis with respect to the normal to the orbital plane
(seasonal variation). For an orbit with an eccentricity $e$, one simply has
$\Delta T/T=e$. In that case, the star to planet distance projected on the sky
varies by an amount $2ea\sin \alpha$ ($= 0.1 a$ in case of an eccentricity
0.1 and $\alpha = 30^o$) between two positions of extreme temperature 
variation.

{\it Diurnal:} The combination of the existence of surface albedo
inhomegeneities (continents and oceans) with the diurnal rotation of the planet
causes a periodic variation of its reflected flux. The period of this 
variation gives the duration of the day and its amplitude gives constraints
on the surface features. For instance, on Earth, the albedo contrast
continents/oceans is 200 to 300 \%, giving rise to a relative flux variation 
of a few tenths for the whole planet. 

{\it Random:} It is due to planet clouds or dust storms. On Earth, the global 
cloud coverage varies from 30\% to 50\%. For a mean surface albedo of 20\% and
 a cloud albedo of 80\%, the resulting random global albedo variation
has an amplitude of 10-20\%.
 
Some of these effects have been estimated quantitatively in the case of the 
Earth (Ford, Turner and Seager 2001). 
Nevertheless, it must be remarked that they will be difficult to 
disentangle from other effects (such as eccentric orbits or binary
planets or planetary rings) without additional information
on the planet position.
 
It is important to point out that 
several conditions must be fulfilled in order to make these effects measurable:
\begin{enumerate}
\item The angular resolution must be sufficient (a fraction of mas)
\item In order to detect both reflected light and thermal emission,
both visible and mid infrared spectral domain must be accessible
\item The photometric precision must be sufficient to detect temperature 
variations (of the order of at most a few percent)
\item It must be possible to make separate exposures shorter than
a planet rotation period at any time along the orbit. This exludes inertial 
scannning modes of the celestial sphere.
\end{enumerate}
\section{Conclusion: constraints on future space mission ``architectures''} 
Imaging is scientifically more productive for
optical architectures having, in addition to obvious angular resolution 
(large baseline) and stellar light rejection power 
\begin{itemize}
\item Sufficiently flexible pointing capabilities
\item Sufficient photometric precision
\item Adequate spectral domain
\end{itemize}
Architectures fulfilling these conditions are feasible. For instance the
``hypertelescope'' concept (Labeyrie 1996) is now under study in the 
framework of the Terrestrial Planet Finder project (Ridgway 2000). 


\begin{thebibliography}{}
\bibitem{DesMarais} Des Marais D., Harwit M., Jucks K., Kasting J., Lunine J.,
Lin D., Seager S., Schneider J., Traub W. and Woolf N., 2002,
Biosignatures and Planetary Properties to be Investigated
by the TPF Mission. JPL Publication 01-008
\bibitem{Ford} Ford E., Seager S. and Turner E. 2001, Characterization of 
extrasolar terrestrial planets from diurnal photometric variability.
    Nature, 412, 885
\bibitem{Labeyrie} Labeyrie A., 1996, Resolved imaging of extra-solar planets
 with future 10-100km optical interferometric arrays.
    Astron. and Astrophys. Suppl., 118, 517
\bibitem{Ridgway}Ridgway S. et al. 2001, 
    Terrestrial Planet Finder: Architectures and Search Strategy. BAAS 32, 
No 4., 70.03  
\bibitem{Schneider99} Schneider J. 1999, The study of extrasolar planets: 
methods of detection, first discoveries and future perspectives.
C.R. Acad. Sci. Paris, 327, Serie II b, 621 
\bibitem{Riaud}Schneider J. and Riaud P., 2002, submitted
\bibitem{Woolf} Woolf, N. 2001 private communication; see also Des Marais et 
al 2002
\end{thebibliography}
\end{document}